\documentclass[12pt]{iopart}
\usepackage{iopams}
 \usepackage{amsthm}
  \usepackage{amssymb}
 \catcode`@=11
\newcommand{\overset}[2]{\binrel@{#2}%
  \binrel@@{\mathop{\kern\z@#2}\limits^{#1}}}
\newcommand{\underset}[2]{\binrel@{#2}%
  \binrel@@{\mathop{\kern\z@#2}\limits_{#1}}}
\catcode`@=12
\begin{document}
\title [Chiral Potts model: Proof of conjecture]%
{Superintegrable chiral Potts model:
Proof of the conjecture for the coefficients of the generating
function ${\mathcal G}(t,u)$}
\author{Helen Au-Yang$^{1}$ and Jacques H H Perk$^{1}$%
\footnote{Supported in part by the National Science Foundation
under grant PHY-07-58139 and by the Australian Research Council
under Project ID DP1096713.}}
\address{$^1$ Department of Physics, Oklahoma State University,
145 Physical Sciences, Stillwater, OK 74078-3072, USA}
\ead{\mailto{perk@okstate.edu}, \mailto{helenperk@yahoo.com}}

\begin{abstract}
In this paper, we prove the conjecture for the coefficients of
the two variable generating function used in our previous paper.
The conjecture was tested numerically before, but its proof was
lacking up to now.
\end{abstract}

\newfont{\mycal}{eufb10 at 12pt}
\newfont{\myeu}{eurm10 at 12pt}
\newcommand{\trace}{{\rm trace}\;}
\newcommand{\Kb}{\overline{K}}
\renewcommand{\i}{{\rm i}}
\newcommand{\half}{{\textstyle\frac{1}{2}}}
\newcommand{\halfs}{{\scriptstyle\frac{1}{2}}}

\renewcommand{\arraystretch}{1.5}
\newcounter{storeeqn}
\renewcommand{\theequation}{\arabic{section}.\arabic{equation}}
\def\vp{^{\vphantom{x}}}
\def\vpt{_{\vphantom{'}}}
\def\bb{{\bar b}}
\def\bI{{\bar{\mathcal I}}}
\def\rd{{\mathrm d}}

\def\sfactor#1#2{\raisebox{-2pt}{$\Bigg[$}{#1\atop#2}
 \raisebox{-2pt}{$\Bigg]$}}
\def\Op{{\underset{\ge}{\Omega}}}
\def\bA{{\bar A}}
\def\bB{{\bar B}}
\def\bC{{\bar C}}
\def\bD{{\bar D}}
\def\bK{{\bar K}}
\def\bN{{\bar N}}
\def\bG{{\bar G}}
\def\bfB{{\bf B}}
\def\bfY{{\bf Y}}
\def\bfU{{\bf U}}
\def\wb{{\overline W}}

\section{Introduction\label{sec1}}
In our previous paper \cite{AuYangPerk2010c} we made a conjecture for
the coefficients of a two-variable generation function, backed by exact
results for special cases and numerical evidence. This enabled us to
derive a formula of Baxter \cite{BaxterII,BaxterIII} for the order
parameter of the chiral Potts model, from which an algebraic
proof \cite{BaxterIII,BaxterIV,BaxterV,Iorgov} of the old conjecture
of Albertini et al.\ \cite{AMPT89} followed. Thus we were able to
obtain the same results by a different method, namely using explicit
expressions for the ground state eigenvectors.

More specifically, the coefficients of the generating function
\begin{eqnarray}
{\mathcal G}(t,u)&=&\sum_{{\{0\le n_j\le N-1\}}\atop{n_1+\cdots+n_L=N}}
\frac{(1-t^N)^{L-1}(1-u^N)^{L-1}}
{\prod_{j=1}^L(1-t\omega^{N_j})(1-u\omega^{-N_j})},\quad
N_j=\sum_{i<j}n_i,
\label{Gtu}\\
&=&\sum_{\ell=0}^{N(r-1)}\sum_{k=0}^{N(r-1)}
{\mathcal G}_{\ell,k}t^\ell u^k,\quad r=(N-1)L/N,
\quad\omega=\e^{2i\pi/N},
\label{gtus}\end{eqnarray}
are symmetric, i.e.\ ${\mathcal G}_{\ell,k}={\mathcal G}_{k,\ell}$,
and for $P\ge Q$ they have been conjectured to be
\begin{eqnarray}
{\mathcal G}_{\ell N+Q,jN+P}={\mathcal G}_{j N+P,\ell N+Q}\nonumber\\
=\sum_{n=0}^j\,\Big[(j-n+1)\Lambda^{Q}_{n}\Lambda^{P}_{\ell+1+j-n}
-(n-\ell)\Lambda^{Q}_{\ell+1+j-n}\Lambda^{P}_{n}\Big],
\label{Id}\end{eqnarray}
where $\Lambda^{P}_{n}=c_{nN+P}$ are the coefficients of the polynomial
\begin{equation}
{\mathcal Q}(t)=
\frac{(1-t^N)^L}{(1-t)^L}=\sum_{m=0}^{L(N-1)} c_m t^m.
\label{Q}\end{equation}
For $P=Q$ this formula has been proven in \cite{AuYangPerk2010a},
using a transformation formula \cite[chapter 10]{AndrewAskeyRoy}
for the hypergeometric series. For $P>Q$, we have shown in
\cite{AuYangPerk2010c}, that there exist messy extra terms which
make it impossible to adopt the $P=Q$ proof of \cite{AuYangPerk2010a}.
 
To calculate the pair correlation of the superintegrable chiral
spin chain, we need to derive identities for a Fourier transform
of the generating function in (\ref{Gtu}). For this reason, we feel
it may be necessary to prove our conjecture using different 
approaches.

As can be seen from the definition of $N_j$ in (\ref{Gtu}),
we have $N_1=0$. The condition $\sum n_i=N$ in (\ref{Gtu}) may
be replaced by $0\le N_2\le N_3\cdots\le N_L\le N$ by excluding
the case $N_i=N$ for all $i=1,\cdots,L$. We rewrite (\ref{Gtu}) as
\begin{equation}
\fl{\mathcal G}(t,u)=\frac{1}{(1\!-\!t^N)(1\!-\!u^N)}\bigg[f_0
\underset{0\le N_2\le N_3\cdots\le N_L\le N}{\sum\sum\cdots\sum}
f_{N_2}f_{N_3}\cdots f_{N_L}-Lf_0^L\bigg],
\label{Gtu2}\end{equation}
where $f_N=f_0$ and
\begin{equation}
f_{N_i}\equiv\frac{(1\!-\!t^N)(1\!-\!u^N)}
{(1\!-\!t\omega^{N_i})(1\!-\!u\omega^{-N_i})}
=\sum_{\mu_i=0}^{N-1}\sum_{\nu_i=0}^{N-1}
t^{\mu_i}u^{\nu_i}\omega^{N_i(\mu_i-\nu_i)}.
\label{fa}\end{equation}

Substituting the above equation into (\ref{Gtu2}), we find
\begin{equation}
{\mathcal G}(t,u)=\frac{1}{(1\!-\!t^N)(1\!-\!u^N)}
\bigg[{\mathcal F}(t,u)-Lf_0^L\bigg],
\label{GF}\end{equation}
where $f_0^L={\mathcal Q}(t){\mathcal Q}(u)$, and 
\begin{equation}
{\mathcal F}(t,u)=\prod_{i=1}^{L}
\sum_{\mu_i=0}^{N-1}\sum_{\nu_i=0}^{N-1}
t^{\mu_i} u^{\nu_i}{\mathcal V}(\{\mu_i\},\{\nu_i\}),
\label{Ftu}\end{equation}
with
\begin{equation}
{\mathcal V}(\{\mu_i\},\{\nu_i\})=
\underset{0\le N_2\le N_3\cdots\le N_L\le N}{\sum\sum\cdots\sum}
\epsilon_2^{N_2}\epsilon_3^{N_3}\cdots\epsilon_L^{N_L},
\quad\epsilon_i=\omega^{\mu_i-\nu_i}.
\label{V}\end{equation}
In section 2, we use the method in \cite[chapter 11]{AndrewAskeyRoy},
to analyze the sum ${\mathcal V}(\{\mu_i\},\{\nu_i\})$, followed by
the analysis of ${\mathcal F}(t,u)$ in sections 3, 4 and 5.

\section{Reduction of the sum ${\mathcal V}(\{\mu_i\},\{\nu_i\})$
by MacMahon method\label{sec2}}
\setcounter{equation}{\value{storeeqn}}
 \renewcommand{\theequation}{\thesection.\arabic{equation}}
We follow an idea of MacMahon described on page 555
in \cite{AndrewAskeyRoy}, to handle the inequality
$0\le N_2\le N_3\cdots\le N_L$, by inserting new variables
$\lambda_2,\lambda_3,\cdots,\lambda_{L-1}$ into (\ref{V}), i.e.
\begin{equation}
\fl{\mathcal V}(\{\mu_i\},\{\nu_i\})=\Op
\sum_{N_2=0}^N\sum_{N_3=0}^N\cdots\sum_{N_L=0}^N\epsilon_2^{N_2}
\epsilon_3^{N_3}\cdots\epsilon_L^{N_L}\lambda_2^{N_3-N_2}
\lambda_3^{N_4-N_3}\cdots\lambda_{L-1}^{N_L-N_{L-1}}.
\label{V1}\end{equation}
By selecting only nonnegative powers of $\lambda_i$, which is denoted
by the operator $\Omega_{\ge}$, sums with the inequality
$0\le N_2\le N_3\cdots\le N_L$ can be replaced by independent sums
over $N_i$ taking all values $0\le N_i\le N$ for $2\le i\le L$.
Even though, we are dealing with the root-of-unity case,
this procedure still works.

Carrying out these sums in (\ref{V1}), we find 
\begin{equation}
\fl{\mathcal V}(\{\mu_i\},\{\nu_i\})=\Op\,
\Bigg[\frac{1-\epsilon_2/\lambda^{N+1}_2}{1-\epsilon_2/\lambda_2}
\cdots\frac{1-\epsilon_j(\lambda_{j-1}/\lambda_j)^{N+1}}
{1-\epsilon_j\lambda_{j-1}/\lambda_j}\cdots
\frac{1-\epsilon_L\lambda^{N+1}_{L-1}}
{1-\epsilon_L\lambda_{L-1}}\Bigg].
\label{V2}\end{equation}
It is easy to combine Lemma 11.2.3 and Proposition 11.3.1
in \cite{AndrewAskeyRoy} in order to obtain
\begin{eqnarray}
\Op\;
&&\frac{\lambda^{-\alpha}}{(1-x\lambda)(1-y_1/\lambda)
(1-y_2/\lambda)\cdots(1-y_M/\lambda)}\nonumber\\
&&=\frac{x^{\alpha}}{(1-x)(1-xy_1)(1-xy_2)\cdots(1-xy_M)},
\qquad\alpha\ge0.
\label{id1}\end{eqnarray}
For the case with positive powers of $\lambda$ in the numerator, the
situation is rather different.
We consider first
\begin{eqnarray}
\Op\;\frac{\lambda^\alpha}{(1-x\lambda)(1-y/\lambda)^n}=
\Op\,\sum_{m=0}^\infty(x\lambda)^m
\sum_{\ell=0}^\infty\frac{(n)_\ell}{\ell!}
y^\ell\lambda^{\alpha-\ell}\nonumber\\
=\sum_{\ell=0}^{\alpha-1}\frac{(n)_\ell}{\ell!}
y^\ell\sum_{m=0}^\infty x^m
+\Op\,\sum_{m=0}^\infty(x\lambda)^m
\sum_{\ell=\alpha}^{\infty}\frac{(n)_\ell}{\ell!}
y^\ell\lambda^{\alpha-\ell}.
\label{idpr}\end{eqnarray}
Replacing $m\to k=m-\ell+\alpha$ in the second term, this becomes
\begin{eqnarray}
\fl\Op\;\frac{\lambda^\alpha}{(1-x\lambda)(1-y/\lambda)^n}
&=&\sum_{\ell=0}^{\alpha-1}\frac{(n)_\ell}{\ell!}y^\ell\frac1{1-x}
+\sum_{k=0}^\infty x^{k-\alpha}\sum_{\ell=\alpha}^{\infty}
\frac{(n)_\ell}{\ell!}(xy)^\ell
\label{id2a}\\
&=&\frac1{1-x}\Bigg[\frac{x^{-\alpha}}{(1-xy)^n}+
\sum_{\ell=0}^{\alpha-1}\frac{(n)_\ell}{\ell!}
y^\ell(1-x^{-\alpha+\ell})\Bigg].
\label{id2}\end{eqnarray}

Let the partial fraction decomposition be
\begin{equation}
\fl\frac1{R(t)}=\sum_{j=1}^k\sum_{i=1}^{n_j}
\frac{a_{j,i}}{(1-{\bar y}_jt)^i},\quad\mbox{where}\quad
R(t)=(1-y_1t)(1-y_2t)\cdots(1-y_Mt).
\label{partial}\end{equation}
Here, for given $j\le k$, $1/{\bar y}_j$ is one of the $k$ distinct
roots of $R(t)$ with multiplicity $n_j$ so that $\sum_{j=1}^k n_j=M$.
We may relate (\ref{partial}) to the symmetric functions
\cite{MacDonald,Sagan} writing
\begin{equation}
\frac1{R(t)}=\sum_{m=0}^\infty S_m t^m,\qquad
S_m=\sum_{1\le a_1\le a_2\cdots\le a_m\le M}
y_{a_1}y_{a_2}\cdots y_{a_m}.
\label{Sm}\end{equation}
Expanding the right-hand side of the first equation in (\ref{partial})
as a series in $t$, and comparing with (\ref{Sm}), we find
\begin{equation}
\fl\frac1{R(t)}=\sum_{j=1}^k\sum_{i=1}^{n_j}{a_{j,i}}
\sum_{m=0}^\infty\frac{(i)_m}{m!}({\bar y}_j t)^m,
\quad\mbox{or}\quad
S_m=\sum_{j=1}^k\sum_{i=1}^{n_j}{a_{j,i}}
\frac{(i)_m}{m!}({\bar y}_j)^m.
\label{partial2}\end{equation}
We are now ready to prove the following proposition:
\newtheorem{Id2}{Proposition}
\begin{Id2}
For $\alpha>0$, we find
\begin{eqnarray}
\fl&&\Op\;\frac{\lambda^{\alpha}}
{(1-x\lambda)(1-y_1/\lambda)(1-y_2/\lambda)\cdots(1-y_M/\lambda)}
=\Op\;\frac{\lambda^{\alpha}}{(1-x\lambda)R(1/\lambda)}
\label{id3a}\\
\fl&&=\frac{x^{-\alpha}}{(1-x)(1-xy_1)(1-xy_2)\cdots(1-xy_M)}
+\sum_{m=0}^{\alpha-1}S_m\frac{(1-x^{m-\alpha})}{1-x}
\label{id3b}\\
\fl&&=\frac 1{1-x}\Bigg[\sum_{m=0}^{\alpha-1}S_m+
\sum_{m=\alpha}^\infty S_m x^{m-\alpha}\Bigg],
\label{id3c}\end{eqnarray}
where $S_m$ is the symmetric function defined in (\ref{Sm}).
\end{Id2}\goodbreak

\begin{proof} 
Substituting (\ref{partial}) into (\ref{id3a}) and using (\ref{id2}),
we find
\begin{equation}
\fl\Op\;\frac{\lambda^{\alpha}}{(1-x\lambda)R(1/\lambda)}
=\frac 1{1-x}\Bigg[\sum_{j=1}^k\sum_{i=1}^{n_j}{a_{j,i}}
\Bigg[\frac{x^{-\alpha}}{(1-x{\bar y}_j)^i}
+\sum_{m=0}^{\alpha-1}
\frac{(i)_m}{m!}{\bar y}_j^m(1-x^{m-\alpha})\Bigg].
\label{pr1}\end{equation}
From (\ref{partial}) and (\ref{partial2}) we obtain (\ref{id3b});
using the first equation in (\ref{Sm}) we get (\ref{id3c}).
This completes the proof.
\end{proof}\goodbreak

Now we define, for $2\le j\le L-1$, 
\begin{eqnarray}
\fl{\mathcal T}_j=\Op\,
\Bigg[\frac{1-(\epsilon_2/\lambda_2)^{N+1}}{1-\epsilon_2/\lambda_2}
\cdot\frac{1-(\lambda_{2}\epsilon_3/\lambda_3)^{N+1}}
{1-\lambda_{2}\epsilon_3/\lambda_3}\cdots
\frac{1-(\lambda_j\epsilon_{j+1}/\lambda_{j+1})^{N+1}}
{1-\lambda_{j}\epsilon_{j+1}/\lambda_{j+1}}\Bigg]\nonumber\\
=\Op\,\bigg[{\mathcal T}_{j-1}
\frac{1-(\lambda_j\epsilon_{j+1}/\lambda_{j+1})^{N+1}}
{1-\lambda_{j}\epsilon_{j+1}/\lambda_{j+1}}\bigg],
\label{Tj}\end{eqnarray}
with $\lambda_1=\lambda_L=1$. Then we use (\ref{id1}) and (\ref{id3c})
to prove by induction the following:

\newtheorem{th1}{Theorem}
\begin{th1}
Let
\begin{eqnarray}
x_j=\epsilon_{j+1}/\lambda_{j+1},\quad y^{(j-1)}_m=
\prod_{i=0}^{m-1}\epsilon_{j-i},
\quad\mbox{for}\quad 1\le m\le j-1;\label{xy}\nonumber\\
R_{j-1}(t)=\prod_{m=1}^{j-1}(1-y^{(j-1)}_m t),
\quad\frac1{R_{j-1}(t)}=\sum_{m=0}^\infty S^{(j-1)}_m t^m.
\label{RSj}\end{eqnarray}
Then
\begin{equation}
\fl{\mathcal T}_j=
\frac 1{(1-x_j)R_{j-1}(x_j)}-\frac{x_j^{N+1}}{1-x_j}
\bigg[\sum_{m=0}^N S^{(j-1)}_m
+\sum_{m=N+1}^\infty S^{(j-1)}_m x_j^{m-N-1}\bigg].
\label{id4}\end{equation}
\end{th1}\goodbreak

\begin{proof} 
For $j=2$, we can see from (\ref{Tj}) that
\begin{equation}
\fl{\mathcal T}_2=\Op\,
\Bigg[\frac{\big(1-(y^{(1)}/\lambda_2)^{N+1}\big)
\big(1-(\lambda_{2}x_2)^{N+1}\big)}{(1-y^{(1)}/\lambda_2)
(1-\lambda_{2}x_2)}\Bigg],
\quad y^{(1)}=\epsilon_{2},\quad x_2=\frac{\epsilon_{3}}{\lambda_{3}}.
\end{equation}
It is straightforward to show by using (\ref{id1}) that
\begin{equation}
\Op\Bigg[\frac{(y^{(1)}/\lambda_2)^{N+1}
\big(1-(\lambda_{2}x_2)^{N+1}\big)}{(1-y^{(1)}/\lambda_2)
(1-\lambda_{2}x_2)}\Bigg]=0.
\end{equation}
Now we use (\ref{id1}) and (\ref{id2a}) for $n=1$ to find
\begin{equation}
\fl{\mathcal T}_2=
\frac1{(1-y^{(1)}x_2)(1-x_2)}-\frac{x^{N+1}_2}{1-x_2}
\Big[\sum_{m=0}^N (y^{(1)})^m
+\sum_{m=N+1}^\infty (y^{(1)})^m x_2^{m-N-1}\Bigg].
\label{T2}\end{equation}
Since $R_1(t)=1-ty^{(1)}$, $S^{(1)}_m=(y^{(1)})^m$,
we have proven (\ref{id4}) for $j=2$.
Next we assume (\ref{id4}) holds for $j$, and prove it
for ${\mathcal T}_{j+1}$. Let
\begin{equation}
y^{(j)}_1=\epsilon_{j+1},\quad
y^{(j)}_{m+1}=\epsilon_{j+1}y^{(j-1)}_m=
\prod_{i=0}^{m}\epsilon_{j-i+1},
\quad x_j=y^{(j)}_1/\lambda,
\label{id4pr1}\end{equation}
where $\lambda=\lambda_{j+1}$, so that
\begin{equation}
(1-x_j)R_{j-1}(x_j)=R_j(1/\lambda)=
\prod_{m=1}^{j}(1-y^{(j)}_{m}/\lambda).
\label{id4pr2}\end{equation}
From the definition in (\ref{Tj}), we find
\begin{equation}
\fl{\mathcal T}_{j+1}=
\Op\,\Bigg[{\mathcal T}_{j}\cdot\frac{1-(\lambda x_{j+1})^{N+1}}
{1-\lambda x_{j+1}}\Bigg]=
\Op\,\Bigg[\frac{1-(\lambda x_{j+1})^{N+1}}
{{\mathcal R}_j(1/\lambda)(1-\lambda x_{j+1})}\Bigg]
-{\mathcal Z}_{j+1}.
\label{Tj1}\end{equation}
where $x_{j+1}=\epsilon_{j+2}/\lambda_{j+2}$ and
\begin{equation}
\fl{\mathcal Z}_{j+1}=
\Op\,\frac{\big(1-(\lambda x_{j+1})^{N+1}\big)
(y^{(j)}_1/\lambda)^{N+1}}
{(1-y^{(j)}_1/\lambda)(1-\lambda x_{j+1})}
\Bigg[\sum_{m=0}^N S^{(j-1)}_m
+\sum_{m=0}^\infty S^{(j-1)}_{m+N+1}
\bigg(\frac{y^{(j)}_1}{\lambda}\bigg)^{m}\Bigg].
\label{z}\end{equation}
Since there are no positive powers of $\lambda$ in the numerator,
we can use (\ref{id1}) to show ${\mathcal Z}_{j+1}=0$. Finally we
use (\ref{id1}) and (\ref{id3c}) to reduce (\ref{Tj1}) to
\begin{equation}
\fl{\mathcal T}_{j+1}=\frac1{R_j(x_{j+1})(1- x_{j+1})}-
\frac{x_{j+1}^{N+1}}{1-x_{j+1}}\Bigg[\sum_{m=0}^{N}S^{(j)}_m
+\sum_{m=N+1}^\infty S^{(j)}_m x_{j+1}^{m-N-1}\Bigg].
\label{Tj2}\end{equation}
This completes the proof.
\end{proof}\goodbreak

Comparing (\ref{V2}) with (\ref{Tj}), and using (\ref{xy})
so that $x_{L-1}=\epsilon_L$, we find from (\ref{id4}) that
\begin{eqnarray}
\fl{\mathcal V}(\{\mu_i\},\{\nu_i\})={\mathcal T}_{L-1}\nonumber\\
\fl=\frac 1{(1-\epsilon_L)R_{L-2}(\epsilon_L)}
-\frac{\epsilon_L^{N+1}}{1-\epsilon_L}\bigg[\sum_{m=0}^N S^{(L-2)}_m
+\sum_{m=N+1}^\infty S^{(L-2)}_m\epsilon_L^{m-N-1}\bigg].
\end{eqnarray}
Using the equivalence of (\ref{id3c}) and (\ref{id3b}),
and dropping the superscripts, we find the result 
\begin{equation}
{\mathcal V}(\{\mu_i\},\{\nu_i\})=
\sum_{m=0}^N S_m\frac{\epsilon^m_L-\epsilon^{N+1}_L}{1-\epsilon_L}
=\sum_{\ell=0}^N\epsilon_L^\ell+
\sum_{m=1}^N S_m\sum_{\ell=0}^{N-m}\epsilon_L^{m+\ell},
\label{Vf}\end{equation}
where $S_0=1$ and further $S_m$ defined in (\ref{Sm}) are now
given by
\begin{eqnarray}
&&S_m=
\underset{1\le a_1\le a_2\cdots\le a_m\le L-2}{\sum\sum\cdots\sum}
y_{a_1}y_{a_2}\cdots y_{a_m}\quad\mbox{with}\nonumber\\
&&y_m=\prod_{i=0}^{m-1}\epsilon_{L-1-i},
\quad\hbox{for}\quad 1\le m\le L-2.
\label{Sy}\end{eqnarray}
Thus we have managed to relate the $L$-fold sum in (\ref{V})
to the $m$-fold sums in (\ref{Sy}) appearing in (\ref{Vf}).

\section{Reduction of ${\mathcal F}(t,u)$\label{sec3}}
\setcounter{equation}{\value{storeeqn}}
 \renewcommand{\theequation}{\thesection.\arabic{equation}}
Define
\begin{equation}
{\mathcal R}_{\ell,m}=\prod_{i=1}^{L}\,
\sum_{\mu_i=0}^{N-1}\sum_{\nu_i=0}^{N-1}\,
t^{\mu_i} u^{\nu_i}\epsilon^\ell_L S_m.
\label{Rlm}\end{equation}
We find from (\ref{Ftu}) and (\ref{Vf}) that
\begin{equation}
{\mathcal F}(t,u)=
\sum_{m=0}^{N}\sum_{\ell=m}^{N}{\mathcal R}_{\ell,m}=
\sum_{\ell=0}^{N}\sum_{m=0}^\ell{\mathcal R}_{\ell,m}.
\label{FR}\end{equation}
Since $S_0=1$, we substitute $\epsilon_i$ defined in (\ref{V}) into
(\ref{Rlm}) and carry out the summations to obtain
\begin{equation}
\fl{\mathcal R}_{\ell,0}= 
\prod_{i=1}^{L-1}\Bigg[\sum_{\mu_i=0}^{N-1}
\sum_{\nu_i=0}^{N-1}t^{\mu_i} u^{\nu_i}\Bigg]
\sum_{\mu_L=0}^{N-1}(t\omega^{\ell})^{\mu_L}
\sum_{\nu_L=0}^{N-1}(u\omega^{-\ell})^{\nu_L}
=f_0^{L-1}f_\ell,
\label{Rl0}\end{equation}
where $f_\ell$ is defined in (\ref{fa}).
Similarly, substituting (\ref{Sy}) into (\ref{Rlm}), and
using $\epsilon_i=\omega^{\mu_i-\nu_i}$, we find
\begin{eqnarray}
\fl{\mathcal R}_{\ell,m}=
\underset{1\le a_1\le a_2\cdots\le a_m\le L-2}{\sum\sum\cdots\sum}
\prod_{i=1}^{L}\,\sum_{\mu_i=0}^{N-1}
\sum_{\nu_i=0}^{N-1}t^{\mu_i} u^{\nu_i}\epsilon^\ell_L
y_{a_1}y_{a_2}\cdots y_{a_m}\nonumber\\
\fl=
\underset{1\le a_1\le a_2\cdots\le a_m\le L-2}{\sum\sum\cdots\sum}
\Bigg[\sum_{\mu_L=0}^{N-1}(\omega^\ell t)^{\mu_L}
\sum_{\nu_L=0}^{N-1}(\omega^{-\ell}u)^{\nu_L}
\prod_{i=L-a_1}^{L-1}\,\sum_{\mu_{i}=0}^{N-1}(\omega^m t)^{\mu_i}
\sum_{\nu_{i}=0}^{N-1}(\omega^{-m}u)^{\nu_i}\nonumber\\
\times\prod_{i=L-a_2}^{L-a_1-1}\,
\sum_{\mu_i=0}^{N-1}(\omega^{m-1} t)^{\mu_i}
\sum_{\nu_i=0}^{N-1}(\omega^{1-m}u)^{\nu_i}\cdots
\prod_{i=1}^{L-a_m-1}\,\sum_{\mu_i=0}^{N-1}t^{\mu_i}
\sum_{\nu_i=0}^{N-1}u^{\nu_i}\Bigg]\nonumber\\
=f_\ell
\underset{1\le a_1\le a_2\cdots\le a_m\le L-2}{\sum\sum\cdots\sum}
f_{m-1}^{a_2-a_1}\cdots f_1^{a_m-a_{m-1}}f_0^{L-1-a_m}.
\label{Rlm1}\end{eqnarray}
As the $a_i$ are now in the exponents, these functions are not
the usual symmetric functions defined in (\ref{Sm}).
 
\subsection{Multiple sum ${\mathcal E}_{m,k}(a,b)$\label{sec3.1}}
Define
\begin{equation}
{\mathcal E}_{m,k}(a,b)=
\underset{a\le a_1\le a_2\cdots\le a_k\le b}{\sum\sum\cdots\sum}
\Bigg(\frac{f_m}{f_{m-1}}\Bigg)^{a_1}
\Bigg(\frac{f_{m-1}}{f_{m-2}}\Bigg)^{a_2}\cdots
\Bigg(\frac{f_{m-k+1}}{f_{m-k}}\Bigg)^{a_k}.
\label{E}\end{equation}
Comparing with (\ref{Rlm1}), we find
\begin{equation}
{\mathcal R}_{\ell,m}=f_0^L\,(f_\ell/f_0)\,{\mathcal E}_{m,m}(1,L-2).
\label{RE}\end{equation}
Letting $a'_i=a_i-1$ for $i=1,\cdots,k$ in (\ref{E}), we can show
\begin{eqnarray}
\fl{\mathcal E}_{m,k}(a,b)=
\underset{a-1\le a'_1\le a'_2\cdots\le a'_k\le b-1}{\sum\sum\cdots\sum}
\Bigg(\frac{f_m}{f_{m-1}}\Bigg)^{a'_1+1}
\Bigg(\frac{f_{m-1}}{f_{m-2}}\Bigg)^{a'_2+1}\cdots
\Bigg(\frac{f_{m-k+1}}{f_{m-k}}\Bigg)^{a'_m+1}\nonumber\\
=\Bigg(\frac{f_{m}}{f_{m-k}}\Bigg){\mathcal E}_{m,k}(a-1,b-1).
\label{E1}\end{eqnarray}
Moreover, using the fact that the sum in (\ref{E}) includes cases
with $a_1=a_2=\cdots=a_n=a$ and 
$a+1\le a_{n+1}\le a_{n+2}\cdots\le a_k\le b$, for $0\le n\le k$,
we find
\begin{equation}
{\mathcal E}_{m,k}(a,b)=\sum_{n=0}^k
\Bigg(\frac{f_m}{f_{m-n}}\Bigg)^{a}{\mathcal E}_{m-n,k-n}(a+1,b).
\label{E2o}\end{equation}
Next, using (\ref{E1}) and replacing $n$ by $k-n$ we rewite
this as
\begin{equation}
{\mathcal E}_{m,k}(a,b)=
\Bigg(\frac{f_m}{f_{m-k}}\Bigg)^{a}
\sum_{n=0}^k{\mathcal E}_{m-k+n,n}(a,b-1).
\label{E2}\end{equation}
From (\ref{RE}) and (\ref{E2}) we then find
\begin{equation}
\sum_{m=0}^\ell{\mathcal R}_{\ell,m} =f_0^L\sum_{m=0}^\ell 
\Bigg(\frac{f_\ell}{f_0}\Bigg){\mathcal E}_{m,m}(1,L-2)=
f_0^L{\mathcal E}_{\ell,\ell}(1,L-1),
\label{E3}\end{equation}
for $0\le\ell\le N$.
Thus from (\ref{FR}) we have
\begin{equation}
{\mathcal F}(t,u)=\sum_{\ell=0}^N\sum_{m=0}^\ell
{\mathcal R}_{\ell,m}=
f_0^L\sum_{\ell=0}^N{\mathcal E}_{\ell,\ell}(1,L-1)
=f_0^L{\mathcal E}_{N,N}(1,L).
\label{FE}\end{equation}

\subsection{Multiple sum ${\mathcal E}_{N,N}(1,L)$\label{sec3.2}}
Noting from (\ref{fa}) that $f_N=f_0$, we find
from (\ref{E}) that 
\begin{equation}
{\mathcal E}_{N,N}(1,L)=
\underset{1\le a_1\le a_2\cdots\le a_N\le L}{\sum\sum\cdots\sum}
f_{N-1}^{a_2-a_1}f_{N-2}^{a_3-a_2}\cdots f_0^{-a_N+a_1},
\label{R0N1}\end{equation}
where the exponents depend on differences of $a_i$'s, so that
one of the summations can be carried out. 
The condition $1\le a_1\le a_2\cdots\le a_N\le L-2$ may be
broken up into the following cases:
\begin{itemize}
\item If $a_1=a_2=\cdots=a_N$ for $1\le a_1\le L$ in (\ref{R0N1}),
the summand becomes $1$ and the contribution of this case to
${\mathcal E}_{N,N}(1,L)$ is ${\mathcal C}_1=L $.
\item When $a_1=\!\cdots\!=a_n<a_{n+1}=\!\cdots\!=a_N$ for
$n=1,\cdots N-1$, we find the contribution to the sum to be
\begin{eqnarray}
{\mathcal C}_2=\sum_{n=1}^{N-1}\sum_{a_n=1}^{L-1}
\sum_{a_{n+1}=a_n+1}^{L}
f_{N-n}^{a_{n+1}-a_n}f_0^{-a_{n+1}+a_n}\nonumber\\
=\sum_{n=1}^{N-1}\sum_{a=1}^{L-1}
\sum_{\alpha=1}^{L-a}f_{N-n}^{\alpha}f_0^{-\alpha}
=\sum_{\alpha=1}^{L-1}(L-\!\alpha)\sum_{n=1}^{N-1}
\Bigg(\frac{f_{N-n}}{f_0}\Bigg)^{\alpha},
\end{eqnarray}
where $\alpha=a_{n+1}-a_n$ and $a=a_n$.
\item When
$a_1\!=\!\cdots\!=\!a_{n_1}\!<\!a_{n_1+1}\!=\!\cdots\!=
\!a_{n_2}\!<\!\cdots\!=\!a_{n_k}\!<\!a_{n_k+1}\!=\!\cdots\!=\!a_N$ 
for $1\le n_1<n_2<\!\cdots\!<n_k\le N-1$, it is straightforward
to show that the contribution to the sum is
\begin{eqnarray}
\fl{\mathcal C}_k=
\underset{1\le n_1<n_2\cdots<n_k\le N-1}{\sum\sum\cdots\sum}
\sum_{a_{n_1}=1}^{L-k}\sum_{a_{n_2}=a_{n_1}+1}^{L-k+1}\cdots
\sum_{a_N=a_{n_k}+1}^{L}f_{N-n_1}^{a_{n_2}-a_{n_1}}\nonumber\\
\times\,f_{N-n_2}^{a_{n_3}-a_{n_2}}
\cdots f_{N-n_k}^{a_N-a_{n_k}}f_0^{a_{n_1}-a_N}.
\end{eqnarray}
Choosing new summation variables $\alpha_i=a_{n_{i+1}}-a_{n_1}$
for $1\le i\le k-1$, $\alpha_k=a_N-a_{n_1}$, and $a=a_{n_1}$,
we can carry out the sum over $a$, ($1\le a\le L-\alpha_k$).
Thus we find
\begin{equation}
{\mathcal C}_k=
\underset{1\le\alpha_1<\alpha_2\cdots<\alpha_k\le L}{\sum\sum\cdots\sum}
(L-\alpha_k){\mathcal Y}_k(\{\alpha_i\}),
\end{equation}
where $(\{\alpha_i\})=(\alpha_1,\alpha_2,\cdots,\alpha_k)$ and
\begin{equation}
\fl{\mathcal Y}_k(\{\alpha_i\})\equiv
\underset{1\le n_1<n_2\cdots<n_k\le N-1}{\sum\sum\cdots\sum}
\Big[\frac{f_{N-n_1}}{f_{N-n_2}}\Big]^{\alpha_1}
\Big[\frac{f_{N-n_2}}{f_{N-n_3}}\Big]^{\alpha_2}
\cdots\Big[\frac{f_{N-n_k}}{f_{0}}\Big]^{\alpha_k}.
\end{equation}
Replacing $N-n_i$ by $n_{k-i+1}$ we may rewrite this as
\begin{equation}
\fl{\mathcal Y}_k={\mathcal Y}_k(\{\alpha_i\})=
\underset{1\le n_1<n_2\cdots<n_k\le N-1}{\sum\sum\cdots\sum}
\Big[\frac{f_{n_k}}{f_{n_{k-1}}}\Big]^{\alpha_1}
\Big[\frac{f_{n_{k-1}}}{f_{n_{k-2}}}\Big]^{\alpha_2}
\cdots\Big[\frac{f_{n_1}}{f_{0}}\Big]^{\alpha_k}.
\label{Yk}\end{equation}
\item 
Similarly, for $a_1\ne a_2\ne\cdots\ne a_N$, we set
$\alpha_i=a_{i}-a_1$ for $1\le i\le N-1$, and carry out
the summation over $a_1$ to find
\begin{equation}
{\mathcal C}_N=
\underset{1\le\alpha_1<\alpha_2<\cdots<\alpha_{N-1}\le L-2}
{\sum\sum\cdots\sum}
(L-\alpha_{N-1}){\mathcal Y}_{N-1},
\end{equation}
where
\begin{equation}
{\mathcal Y}_{N-1}=\Big[\frac{f_{N-1}}{f_{N-2}}\Big]^{\alpha_1}
\Big[\frac{f_{N-2}}{f_{N-3}}\Big]^{\alpha_2}
\cdots\Big[\frac{f_{1}}{f_{0}}\Big]^{\alpha_k}.
\label{YN}\end{equation}
This case is essentially the previous case with $a_{n_i}\equiv a_i$.
\end{itemize}
Adding all the cases, we find from (\ref{FE}) 
\begin{eqnarray}
\fl{\mathcal F}(t,u)=f_0^L{\mathcal E}_{N,N}(1,L)=
f_0^L\Bigg[L+\sum_{\alpha=1}^{L}(L-\alpha){\mathcal Y}_1+
\underset{1\le\alpha_1<\alpha_2\le L}{\sum\sum}
(L-\alpha_2){\mathcal Y}_2\nonumber\\
+\cdots+
\underset{1\le\alpha_1<\alpha_2<\cdots<\alpha_{N-1}\le L-2}
{\sum\sum\cdots\sum}
(L\!-\alpha_{N-1}){\mathcal Y}_{N-1}\Bigg].
\label{ENN}\end{eqnarray}

\section{Further simplification of ${\mathcal F}(t,u)$
and ${\mathcal G}(t,u)$\label{sec4}}
\setcounter{equation}{\value{storeeqn}}
 \renewcommand{\theequation}{\thesection.\arabic{equation}}
Let
\begin{equation}
\fl{\mathcal J}_k\equiv f_0^L
\underset{1\le a_1<a_2<\cdots<a_{k}\le L-1}{\sum\sum\cdots\sum}
(L\!-\!a_{k}){\mathcal Y}_{k},\qquad
{\mathcal F}(t,u)=Lf_0^L+\sum_{k=1}^{N-1}{\mathcal J}_k.
\label{Jk}\end{equation}
where ${\mathcal Y}_{k}$ is defined in (\ref{Yk}). 
Using induction we now prove the following identity for partial
$\ell$-fold sums appearing in (\ref{ENN}) with (\ref{Yk}) substituted:

\newtheorem{Idsum}{Proposition}
\begin{Idsum}
For $\ell\le k$, we have
\begin{eqnarray}
{\mathcal U}_\ell(a,b)\equiv
\underset{a\le a_1<a_2\cdots<a_\ell\le b}{\sum\sum\cdots\sum}
\Bigg[\frac{f_{n_k}}{f_{n_{k-1}}}\Bigg]^{a_1}
\Bigg[\frac{f_{n_{k-1}}}{f_{n_{k-2}}}\Bigg]^{a_2}
\cdots\Bigg[\frac{f_{n_{k-\ell+1}}}{f_{n_{k-\ell}}}\Bigg]^{a_\ell}
\nonumber\\
\hspace{1in}=\sum_{i=0}^{\ell}
\frac{(f_{n_{k-i}}/f_{n_{k-\ell}})^{b+1}(f_{n_{k}}/f_{n_{k-i}})^{a-1}}
{\prod_{j=0,j\ne i}^{\ell}(f_{n_{k-i}}/f_{n_{k-j}}-1)}
\label{id4a}\end{eqnarray}
\end{Idsum}\goodbreak

\begin{proof} 
Consider $\ell=1$. Then
\begin{equation}
{\mathcal U}_1(a,b)=
\sum_{a_1=a}^{b}\Bigg[\frac{f_{n_k}}{f_{n_{k-1}}}\Big]^{a_1}
=\frac{(f_{n_{k}}/f_{n_{k-1}})^{a}}{1-(f_{n_{k}}/f_{n_{k-1}})}
+\frac{(f_{n_{k}}/f_{n_{k-1}})^{b+1}}{(f_{n_{k-1}}/f_{n_{k}})-1},
\label{id4b}
\end{equation}
showing that (\ref{id4a}) holds for $\ell=1$.

Assuming (\ref{id4a}) holds for $\ell$, we shall show it also
holds for $\ell+1$. Defining $g_i$ via the partial fraction
decomposition 
\begin{equation}
\fl\frac{z^m}{\prod_{i=0}^{\ell}(z/f_{n_{k-i}}-1)}
=\sum_{i=0}^{\ell}\frac{g_i f_{n_{k-i}}^m}{z/f_{n_{k-i}}-1},\quad
g_i\equiv
\frac 1{\prod_{j=0,j\ne i}^{\ell}(f_{n_{k-i}}/f_{n_{k-j}}-1)},
\label{pfrac}\end{equation}
for $m\le\ell$, we can rewrite (\ref{id4a}), valid for $\ell$, as
\begin{equation}
{\mathcal U}_{\ell}(a,a_{\ell+1}\!-\!1)=\sum_{i=0}^{\ell}g_i
\Bigg(\frac{f_{n_{k}}}{f_{n_{k-i}}}\Bigg)^{a-1}
\Bigg(\frac{f_{n_{k-i}}}{f_{n_{k-\ell}}}\Bigg)^{a_{\ell+1}}. 
\label{id4c}
\end{equation}
We then observe that ${\mathcal U}_{\ell}(a,a_{\ell+1}\!-\!1)=0$
for $a\le a_{\ell+1}<a+\ell$,
because $\sum_{i=0}^{\ell}g_i f_{n_{k-i}}^m=0$ for $0<m\le\ell$.
Therefore, we may extend the interval of summation over $a_{\ell+1}$
and find
\begin{eqnarray}
\fl{\mathcal U}_{\ell+1}(a,b)=
\sum_{a_{\ell+1}=a}^{b}{\mathcal U}_\ell(a,a_{\ell+1}\!-\!1)
\Bigg(\frac{f_{n_{k-\ell}}}{f_{n_{k-\ell-1}}}\Bigg)^{a_{\ell+1}}
\nonumber\\
=\sum_{i=0}^{\ell}g_i\sum_{a_{\ell+1}=a}^{b}
\Bigg(\frac{f_{n_{k}}}{f_{n_{k-i}}}\Bigg)^{a-1}
\Bigg(\frac{f_{n_{k-i}}}{f_{n_{k-\ell}}}\Bigg)^{a_{\ell+1}}
\Bigg(\frac{f_{n_{k-\ell}}}{f_{n_{k-\ell-1}}}\Bigg)^{a_{\ell+1}}.
\label{id4d}\end{eqnarray}
Carrying out the summation over $a_{\ell+1}$, we find
\begin{equation}
\fl{\mathcal U}_{\ell+1}(a,b)
=\sum_{i=0}^{\ell}g_i
\frac{(f_{n_{k-i}}/f_{n_{k-\ell-1}})^a-
(f_{n_{k-i}}/f_{n_{k-\ell-1}})^{b+1}}
{1-f_{n_{k-i}}/f_{n_{k-\ell-1}}}
\Bigg(\frac{f_{n_{k}}}{f_{n_{k-i}}}\Bigg)^{a-1}.
\label{id4e}\end{equation}
The $b$-independent part becomes the term $i=\ell+1$ in
(\ref{id4a}) for ${\mathcal U}_{\ell+1}(a,b)$, as
\begin{equation}
\sum_{i=0}^{\ell}\frac{g_i}{f_{n_{k-\ell-1}}/f_{n_{k-i}}-1}=
\frac1{\prod_{i=0}^{\ell}(f_{n_{k-\ell-1}}/f_{n_{k-i}}-1)}
\end{equation}
follows from (\ref{pfrac}). The terms $i=0,\ldots,\ell$
are obtained rewriting 
\begin{equation}
-\frac{g_i}{1- f_{n_{k-i}}/f_{n_{k-\ell-1}}}=
\frac 1{\prod_{j=0,j\ne i}^{\ell+1}(f_{n_{k-i}}/f_{n_{k-j}}-1)}.
\end{equation}
This shows the identity in (\ref{id4a}) holds for $\ell+1$,
thus completing the proof.
\end{proof}\goodbreak
We have not found a general formula, valid for all $\ell$,
if in (\ref{id4a}), the condition $1\le a_1<a_2\cdots<a_\ell$
is replaced by $1\le a_1\le a_2\cdots\le a_\ell$. Therefore,
the breaking up of the sum in subsection \ref{sec3.2} has been
necessary.

\subsection{${\mathcal J}_k$\label{sec4.1}}
From (\ref{Jk}) and (\ref{Yk}), we find
\begin{eqnarray}
\fl{\mathcal J}_k=f_0^L
\underset{1\le a_1<a_2<\cdots<a_{k}\le L}{\sum\sum\cdots\sum}
(L-a_k)\nonumber\\
\times
\underset{1\le n_1<n_2<\cdots<n_k\le N-1}{\sum\sum\cdots\sum}
\Bigg(\frac{f_{n_k}}{f_{n_{k-1}}}\Bigg)^{a_1}
\cdots\Bigg(\frac{f_{n_2}}{f_{n_1}}\Bigg)^{a_{k-1}}
\Bigg(\frac{f_{n_1}}{f_0}\Bigg)^{a_{k}}.
\end{eqnarray}
Using (\ref{id4a}) with $\ell=k-1$, $a=1$, and $b=a_k-1$
to carry out the $(k-1)$-fold sum over $a_1,\ldots,a_{k-1}$,
and using the identity
\begin{equation}
\sum_{a=1}^{L}(L-a)x^a=x^L\sum_{a=1}^{L-1}a x^{-a}=
\frac{x(x^L-1)}{(x-1)^2}-\frac{Lx}{x-1},
\end{equation}
to carry out the sum over $a_k$, we find
\begin{eqnarray}
\fl{\mathcal J}_k=f_0^L
\underset{1\le n_1<n_2<\cdots<n_k\le N-1}{\sum\sum\cdots\sum}
\sum_{a_k=1}^{L-1}(L-a_k)
\sum_{i=1}^{k}\frac{(f_{n_i}/f_0)^{a_k}}
{\prod_{j=1,j\ne i}^{k}(f_{n_i}/f_{n_j}-1)}\nonumber\\
\fl=
\underset{1\le n_1<n_2<\cdots<n_k\le N-1}{\sum\sum\cdots\sum}
\sum_{i=1}^{k}\frac{(f_{n_i}/f_0)}{\prod_{j=1,j\ne i}^{k}
(f_{n_i}/f_{n_j}-1)(f_{n_i}/f_0-1)}
\bigg[\frac{f_{n_i}^L-f_0^L}{(f_{n_i}/f_0-1)}-Lf_0^L\bigg]
\nonumber\\
=\sum_{\mathrm{all}\,\Omega_k}\sum_{r\in\Omega_k}
\frac{(f_{r}/f_0)\prod_{s\in{\bar\Omega}_k}(f_r/f_s-1)}
{\prod_{s=0,s\ne r}^{N-1}(f_{r}/f_s-1)}
\bigg[\frac{f_{r}^L-f_0^L}{(f_{r}/f_0-1)}-Lf_0^L\bigg].
\label{Jk2}\end{eqnarray}
where we let $r=n_i$, so that $1\le i\le k$ becomes
$r\in\Omega_k$, in which $\Omega_k=\{n_1,n_2,\cdots,n_k\}$.
In the last step we multiplied the numerator and the
denominator by the product over the complement of $\Omega_k$,
given by ${\bar\Omega}_{k}=\{1\le s\le N-1|s\notin\Omega_k\}$.
We now let
\begin{equation}
A_r=(1-t\omega^r)(1-u\omega^{-r}),\quad\hbox{so that}
\quad f_r/f_s-1=(A_s-A_r)/A_r.
\label{Ar}\end{equation}
with $f_r$ defined in (\ref{fa}). It is easy to evaluate
\begin{equation}
\fl\prod_{s=0,s\ne r}^{N-1}(A_s-A_r)=
\prod_{s=0,s\ne r}^{N-1}(\omega^r-\omega^s)(t-\omega^{-r-s}u)
=N(t^N-u^N)/(\omega^r t-\omega^{-r}u).
\end{equation}
Consequently, if we let
\begin{equation}
h(r)\equiv\frac{(\omega^r t-\omega^{-r}u)A_0}{N(t^N-u^N)}
\Bigg[\frac{f_r^L-f_0^L}{(A_0/A_r-1)}-Lf_0^L\Bigg],
\label{hr}\end{equation}
we may rewrite (\ref{Jk2}) as
\begin{equation}
\fl{\mathcal J}_k=\sum_{\mathrm{all}\,\Omega_k}\sum_{r\in\Omega_k}
A_r^{k-1}h(r)\prod_{s\in{\bar\Omega}_k}(A_s-A_r)=
\sum_{\mathrm{all}\,{\bar\Omega}_k}\sum_{r=1}^{N-1}
A_r^{k-1}h(r)\prod_{s\in{\bar\Omega}_k}(A_s-A_r),
\label{Jk3}\end{equation}
where we replaced the sum over $r\in\Omega_k$ by the sum over
the entire set $\Omega_k\cup{\bar\Omega}_k$, for the product
over $s$ is identically zero if $r\in{\bar\Omega}_k$;
we also replaced the sum over all sets $\Omega_k$ by the sum
over all ${\bar\Omega}_k$, as these are in one-to-one
correspondence. Consider now the polynomial
\cite{MacDonald,Sagan}
\begin{equation}
\fl\prod_{\ell=1}^{N-1}\big[z+(A_\ell-A_r)\big]=
\sum_{\ell=0}^{N-2}z^{N-1-\ell}e_\ell,\qquad
e_\ell\equiv
\underset{1\le n_1<n_2<\cdots<n_\ell\le N-1}{\sum\sum\cdots\sum}
\prod_{i=1}^\ell (A_{n_i}-A_r).
\label{poly}\end{equation}
In (\ref{Jk3}), the set $\Omega_k$ has $k$ elements;
its complement ${\bar\Omega}_k$ has $N-1-k$ elements $\bar n_i$
with $1\le{\bar n}_1<{\bar n}_2<\cdots<{\bar n}_{N-1-k}$.
Therefore, the sum over ${\bar\Omega}_k$ in (\ref{Jk3}) leads to
$e_{N-1-k}$, or more precisely
\begin{equation}
\fl{\mathcal J}_k=
\sum_{r=1}^{N-1}A_r^{k-1}e_{N-1-k}\,h(r),
\quad\hbox{and}\quad
\sum_{k=1}^{N-1}{\mathcal J}_k=\sum_{r=1}^{N-1}
\sum_{k=1}^{N-1} A_r^{k-1}e_{N-1-k}\,h(r).
\label{Jk4}\end{equation}
From (\ref{poly}) with $z=A_r$, we find
\begin{equation}
\fl\sum_{k=1}^{N-1} A_r^{k-1}e_{N-1-k}=
{A_r}^{-1}\prod_{\ell=1}^{N-1}(A_r+A_\ell-A_r)=
{A_r}^{-1}\prod_{\ell=1}^{N-1}A_\ell={f_0}/{A_r},
\label{Jk5}\end{equation}
where (\ref{Ar}) and (\ref{fa}) have been used.
Applying (\ref{GF}), (\ref{Jk}), (\ref{hr}), (\ref{Jk4}),
(\ref{Jk5}) and (\ref{fa}) in this order, we obtain
\begin{equation}
{\mathcal G}(t,u)=
\frac1{N(t^N-u^N)}\sum_{r=1}^{N-1}(t\omega^r-u\omega^{-r})
\Bigg[\frac{f_r^L-f_0^L}{A_0-A_r}-\frac{Lf_0^L}{A_r}\Bigg].
\label{Gtuf}\end{equation}

\section{Final result for the sum ${\mathcal G}(t,u)$\label{sec5}}
\setcounter{equation}{\value{storeeqn}}
 \renewcommand{\theequation}{\thesection.\arabic{equation}}

\subsection{Useful identities\label{sec5.1}}
It is straightforward to show for $k\ge0$ that
\begin{equation}
\fl\sum_{a=0}^{N-1}\frac{\omega^{-ak}}{1-z\omega^a}
=\sum_{n=0}^\infty z^n\sum_{a=0}^{N-1}\omega^{an-ak}
=N\sum_{n=0}^\infty z^n\sum_{p=0}^\infty\delta_{n,k+pN}
=\frac{N z^k}{1-z^N}.
\end{equation}
Consequently, by substracting the 0th term, we find
\begin{equation}
\sum_{a=1}^{N-1}\frac{\omega^{-ak}}{1-z\omega^a}=
\frac{N z^k}{1-z^N}-\frac 1{1-z}.
\label{sum1}\end{equation}
Taking the limit $z\to 1$, we get
\begin{equation}
\sum_{a=1}^{N-1}\frac{\omega^{-ak}}{1-\omega^a}=\frac12(N -1-2k).
\label{sum2}\end{equation}
If $-N\le k<0$, we have to replcace $k$ by $k+N$ in (\ref{sum1})
and (\ref{sum2}).

\subsection{The sums ${\mathcal B}_1$ and
${\mathcal B}_2$\label{sec5.2}}
We can split the sum in (\ref{Gtuf}), putting the contributions
from the two terms within the square brackets into separate sums,
writing
\begin{equation}
{\mathcal G}(t,u)=({\mathcal B}_1-{\mathcal B}_2)\big/[{N(t^N-u^N)}].
\label{GB1B2}\end{equation}
Partial fraction decomposition yields
\begin{equation}
\frac{t\omega^r-u\omega^{-r}}{A_r}=
\frac{t\omega^r-u\omega^{-r}}{(1-t\omega^{r})(1-u\omega^{-r})}
=\frac 1{1-t\omega^{r}}-\frac 1{1-u\omega^{-r}}.
\end{equation}
Hence, we can use (\ref{sum1}) with $k=0$ to evaluate
${\mathcal B}_2$ as
\begin{equation}
\fl{\mathcal B}_2\equiv
Lf_0^L\sum_{r=1}^{N-1}\frac{t\omega^r-u\omega^{-r}}{A_r}
=Lf_0^L\Bigg[\frac{N(t^N-u^N)}
{(1-t^N)(1-u^N)}-\frac{t-u}{(1-t)(1-u)}\Bigg],
\label{B2}\end{equation}
We may also decompose
\begin{equation}
\frac{t\omega^r-u\omega^{-r}}{A_0-A_r}=
\frac t{t-u\omega^{-r}}+\frac 1{\omega^r-1}.
\label{pfrac2}\end{equation}
We note $f_r^L={\mathcal Q}(t\omega^r){\mathcal Q}(u\omega^{-r})$,
see (\ref{Q}) and (\ref{fa}). Therefore, we may write
\begin{equation}
f_r^L-f_0^L=\sum_{Q=0}^{N-1}\sum_{\ell=0}^{m_Q}
\sum_{P=0}^{N-1}\sum_{m=0}^{m_P}
t^{\ell N+Q}u^{m N+P}\Lambda^Q_\ell\Lambda^P_m
(\omega^{r(Q-P)}-1),
\label{QQ}\end{equation}
where $m_Q\equiv\lfloor(N\!-\!1)L/N-Q/N\rfloor$ and a similar
expression with $Q$ replaced by $P$, first introduced as
(2.17) in \cite{AMPT89}. Therefore,
from (\ref{pfrac2}) and (\ref{QQ}), we find that
\begin{eqnarray}
\fl{\mathcal B}_1\equiv\sum_{r=1}^{N-1}(f_r^L-f_0^L)
\Bigg[\frac{t\omega^r-u\omega^{-r}}{A_0-A_r}\Bigg]\nonumber\\
\fl=\sum_{Q=0}^{N-1}
\sum_{\ell=0}^{m_Q}\sum_{P=0}^{N-1}\sum_{m=0}^{m_P}
t^{\ell N+Q}u^{m N+P}\Lambda^Q_\ell\Lambda^P_m
\sum_{r=1}^{N-1}(\omega^{r(Q-P)}-1)\Bigg[\frac t{t-u\omega^{-r}}+
\frac1{\omega^r-1}\Bigg].
\label{B1a}\end{eqnarray}
From (\ref{sum1}), we can easily show
\begin{eqnarray}
\fl\sum_{r=1}^{N-1}(\omega^{r(Q-P)}-1)
\Bigg[\frac t{t-u\omega^{-r}}\Bigg]
=\begin{cases}
{\frac{N\big[(t/u)^{P-Q}-(t/u)^N\big]}{(t/u)^N-1},&$P>Q$,\\
\frac{N\big[(t/u)^{P-Q+N}-(t/u)^N\big]}{(t/u)^N-1},&$P\le Q$.}
\end{cases}
\label{case1}\end{eqnarray}
Using (\ref{sum2}), we find
\begin{eqnarray}
\sum_{r=1}^{N-1}\frac{\omega^{r(Q-P)}-1}{\omega^r-1}
=\begin{cases}{{P-Q},&$P\ge Q$,\\
{P-Q+N},&$P<Q$.}
\end{cases}
\label{case2}\end{eqnarray}
The sums in (\ref{B1a}), (\ref{case1}) and (\ref{case2}) are all
identically zero for $P=Q$. Therefore, using (\ref{case1}) and
(\ref{case2}), we can rewrite (\ref{B1a}) as
\begin{eqnarray}
\fl{\mathcal B}_1=\sum_{\ell}\sum_{m}\Bigg\{\sum_{P>Q}
t^{\ell N+Q}u^{m N+P}\Lambda^Q_\ell\Lambda^P_m
\Bigg[\frac{N\big[(t/u)^{P-Q}-(t/u)^N\big]}{(t/u)^N-1}+P-Q\Bigg]
\nonumber\\
+\sum_{P<Q}t^{\ell N+Q}u^{m N+P}\Lambda^Q_\ell\Lambda^P_m
\Bigg[\frac{N\big[(t/u)^{P-Q+N}-1\big]}
{(t/u)^N-1}+P-Q\Bigg]\Bigg\}.
\label{B1b}\end{eqnarray}
We may split ${\mathcal B}_1$ into two parts, one having the factor $P-Q$
\begin{eqnarray}
\fl{\mathcal B}_1=\beta_1+\beta_2,\quad
{\beta}_2&\equiv&
\sum_{\ell}\sum_{m}\Bigg[\sum_{P>Q}+\sum_{P<Q}\Bigg]
\Big[t^{\ell N+Q}u^{m N+P}\Lambda^Q_\ell\Lambda^P_m(P-Q)\Big]
\nonumber\\
&=&\sum_{\ell}\sum_{m}\sum_{P}\sum_{Q}
\Big[t^{\ell N+Q}u^{m N+P}\Lambda^Q_\ell\Lambda^P_m(P-Q)\Big],
\label{beta2}\end{eqnarray}
In the other part of (\ref{B1b}), we interchange
$m\leftrightarrow\ell$ and $P\leftrightarrow Q$ in the sums
with $P<Q$ to find
\begin{equation}
\fl{\beta}_1=\frac N{t^N-u^N}
\sum_{\ell}\sum_{m}\sum_{P>Q}
\Lambda^Q_\ell\Lambda^P_m
(t^{\ell N}u^{mN}-t^{mN}u^{\ell N})(t^Pu^{N+Q}-t^{N+Q}u^P).
\label{beta1}\end{equation}
In (\ref{beta2}), we let $r=\ell N+Q$ and $s=mN+P$ so that
$c_r=\Lambda^Q_\ell$, $c_s=\Lambda^P_m$ which are the coeffients
of the polynomial in (\ref{Q}). 
Because $P-Q=s-r+\ell N-m N$, we find
\begin{equation}
\fl{\beta}_2=
\sum_{r}\sum_{s}(s-r)c_r c_s t^r u^s+
N\sum_{\ell}\sum_{m}\sum_{P}\sum_{Q}(\ell-m)
\Big[t^{\ell N+Q}u^{m N+P}\Lambda^Q_\ell\Lambda^P_m\Big].
\label{beta2a}\end{equation}
The first term is denoted by $\gamma_1$ and second by $\gamma_2$,
so that
\begin{equation}
\fl\beta_2=\gamma_1+\gamma_2,
\quad
{\gamma}_1\equiv\sum_{r}\sum_{s}(s-r)c_r c_s t^r u^s=
u{\mathcal Q}(t){\mathcal Q'}(u)-t{\mathcal Q'}(t){\mathcal Q}(u).
\end{equation}
Using $f_0^L={\mathcal Q}(t){\mathcal Q}(u)$, we may also write
\begin{equation}
\fl{\gamma}_1=f_0^L\Big[u\frac{d\ln{\mathcal Q}(u)}{du}-
t\frac{d\ln{\mathcal Q}(t)}{dt}\Big]
=Lf_0^L\Bigg[\frac{N(t^N-u^N)}{(1-t^N)(1-u^N)}-
\frac{t-u}{(1-t)(1-u)}\Bigg],
\label{gamma1}\end{equation}
which is identical to ${\mathcal B}_2$ in (\ref{B2}).
We split the summation over $P$ and $Q$ of the second term $\gamma_2$
in (\ref{beta2a}) into $P=Q$, $P>Q$ and $P<Q$. For $P<Q$, we 
interchange the variables $P\leftrightarrow Q$ and
$m\leftrightarrow\ell$ to find 
\begin{eqnarray}
\gamma_2=\psi_1+\psi_2,\quad 
\psi_1\equiv N\sum_{\ell}\sum_{m}\sum_{P=0}^{N-1}(\ell-m)
t^{\ell N+P}u^{m N+P}\Lambda^P_\ell\Lambda^P_m\label{psi1},\\
\psi_2= N\sum_{\ell}\sum_{m}\sum_{P>Q}(\ell-m)\Lambda^Q_\ell\Lambda^P_m
(t^{\ell N+Q}u^{m N+P}-t^{m N+P}u^{\ell N+Q}).
\label{psi2}\end{eqnarray}
Since $\gamma_1={\mathcal B}_2$, we find
\begin{equation}
{\mathcal B}_1-{\mathcal B}_2=\beta_1+\gamma_1+\gamma_2-{\mathcal B}_2
=\beta_1+\gamma_2=\beta_1+\psi_1+\psi_2,
\end{equation}
so that (\ref{GB1B2}) becomes
\begin{equation}
{\mathcal G}(t,u)=({\beta_1+\psi_1+\psi_2})\big/[{N(t^N-u^N)}].
\label{Gtufi}\end{equation}

\subsection{The sum $\psi_1$\label{sec5.3}}
We now split the summation over $\ell$ and $m$ in (\ref{psi1})
into three parts: $\ell>m$, $\ell<m$ and $\ell=m$. The summand
is identically zero for $\ell=m$ and combining the other two parts
by interchanging the variables $\ell\leftrightarrow m$ for the part
with $\ell<m$, we obtain
\begin{eqnarray}
\fl\psi_1=N\sum_{P=0}^{N-1}\sum_{\ell>m}(\ell-m)
\Lambda^P_\ell\Lambda^P_m (tu)^P(t^{\ell N}u^{m N}- t^{m N}u^{\ell N})
\nonumber\\
\fl=N(t^N-u^N)\sum_{P=0}^{N-1}\sum_{\ell=1}^{m_P}
\sum_{m=0}^{\ell-1}(\ell-m)\Lambda^P_\ell\Lambda^P_m (tu)^{P+mN}
\sum_{s=0}^{\ell-m-1}t^{(\ell-m-1-s) N}u^{s N},
\label{psi1a}\end{eqnarray}
where $m_P=\lfloor(N-1)L/N-P/N\rfloor$. We need to change the
summation variables $m$ and $\ell$, so that we can read off
the coefficients of $t^{\ell N+P}u^{mN+P}$. To do so, we first
interchange the order of summations to move the sum over $s$ to
the left, followed by the change of summation variables to
$\ell'=\ell-1-s$ and $m'=m+s$ and the move of the summation over
$s$ back to the right. More explicitly,
\begin{eqnarray}
\fl\psi_1'\equiv\frac{\psi_1}{N(t^N-u^N)}=
\sum_{P=0}^{N-1}\sum_{s=0}^{m_P-1}
\sum_{\ell=s+1}^{m_P}\sum_{m=0}^{\ell-1-s}(\ell-m)
\Lambda^P_\ell\Lambda^P_m t^{(\ell-1-s) N+P}u^{(m+s)N+P}
\label{psi1b}\\
\fl=\sum_{P=0}^{N-1}\sum_{s=0}^{m_P-1}
\sum_{\ell'=0}^{m_P-s-1}\sum_{m'=s}^{\ell'+s}(\ell'\!-\!m'+2s+1)
\Lambda^P_{\ell'+1+s}\Lambda^P_{m'-s} t^{\ell'N+P}u^{m'N+P}
\label{psi1c}\\
\fl=\sum_{P=0}^{N-1}\sum_{\ell=0}^{m_P-1}
\sum_{m=0}^{m_P-1}t^{\ell N+P}u^{mN+P}
\sum_{s=\max(0,m-\ell)}^{m}(\ell-m+2s+1)
\Lambda^P_{\ell+1+s}\Lambda^P_{m-s}, 
\label{psi1d}\end{eqnarray}
dropping the primes on $\ell$ and $m$ in the last step.
In (\ref{psi1c}) we have $0\le s\le m'\le\ell'+s$, implying 
$\max(0,m-\ell)\le s\le m$ in (\ref{psi1d}).

It is easily seen that $\beta_1$ and $\psi_2$ vanish if $P=Q$.
Therefore, from (\ref{Gtufi}) we conclude that the coefficient
of $t^{\ell N+P}u^{mN+P}$ in (\ref{psi1d}) gives
\begin{equation}
{\mathcal G}_{\ell N+P,m N+P}=
\sum_{s=\max(0,m-\ell)}^{m}(\ell-m+2s+1)
\Lambda^P_{\ell+1+s}\Lambda^P_{m-s},
\label{GPP1}\end{equation}
for all $0\le\ell,m\le m_P-1$. We note that (\ref{GPP1})
is identical to equation (37) in \cite{AuYangPerk2010a}
for $m\le\ell$. Making the substitution $s\to m-s$ we find
the alternative expression
\begin{equation}
{\mathcal G}_{\ell N+P,m N+P}=\sum_{s=0}^{\min(m,\ell)}(\ell+m-2s+1)
\Lambda^P_{\ell+m+1-s}\Lambda^P_{s},
\label{GPP2}\end{equation}
agreeing with (44) of \cite{AuYangPerk2010a} for $m\le\ell$.
Expression (\ref{GPP2}), and thus also (\ref{GPP1}), is symmetric
in $\ell$ and $m$, as could be expected from the original
definitions (\ref{Gtu}) and (\ref{gtus}).

\subsection{Evaluation of the sum $\psi_2+\beta_1$\label{sec5.4}}
From (\ref{beta1}) and (\ref{psi2}) we find that the summands are
identically zero for $m=\ell$. The summation over $m$ and $\ell$ can be
split into the two contributions $\ell>m$ and $\ell<m$, resulting in
\begin{equation}
\psi_2+\beta_1=\alpha_1+\alpha_2,
\label{alpha12}\end{equation}
where
\begin{eqnarray}
\fl\alpha_1\equiv N\sum_{P>Q}\sum_{\ell>m}\Lambda^Q_\ell\Lambda^P_m
\Bigg[(\ell-m)(t^{\ell N+Q}u^{mN+P}-t^{mN+P}u^{\ell N+Q})
\nonumber\\
+(t^P u^{N+Q}-t^{N+Q}u^P)
\sum_{s=0}^{\ell-m-1}t^{(\ell-1-s)N}u^{(m+s)N}
\Bigg],
\label{alpha1}\\
\fl\alpha_2\equiv N\sum_{P>Q}\sum_{\ell<m}\Lambda^Q_\ell\Lambda^P_m
\Bigg[(\ell-m)(t^{\ell N+Q}u^{mN+P}-t^{mN+P}u^{\ell N+Q})
\nonumber\\
-(t^P u^{N+Q}-t^{N+Q}u^P)
\sum_{s=0}^{m-\ell-1}t^{(m-1-s)N}u^{(\ell+s)N}\Bigg].
\label{alpha2}\end{eqnarray}
Using
\begin{equation}
\fl\sum_{s=0}^{\ell-m-1}1=\ell-m,\quad 
\sum_{s=0}^{\ell-m-1}t^{(\ell-1-s)N}u^{(m+s)N}
=\sum_{s=0}^{\ell-m-1}t^{(m+s)N}u^{(\ell-1-s)N},
\end{equation}
we may rewrite (\ref{alpha1}) as
\begin{eqnarray}
\fl\alpha_1=N\sum_{P>Q}\sum_{\ell>m}
\Lambda^Q_\ell\Lambda^P_m\sum_{s=0}^{\ell-m-1}
(t^{\ell N+Q}u^{mN+P}-t^{mN+P}u^{\ell N+Q}
\nonumber\\
+t^{(m+s)N+P}u^{(\ell-s)N+Q}-t^{(\ell-s)N+Q}u^{(m+s)N+P})
\nonumber\\
\fl=N\sum_{P>Q}\sum_{\ell>m}
\Lambda^Q_\ell\Lambda^P_m\sum_{s=0}^{\ell-m-1}
(t^{(\ell-s)N+Q}u^{mN+P}+t^{mN+P}u^{(\ell-s)N+Q})(t^{sN}-u^{sN})
\nonumber\\
\fl=N(t^N-u^N)\sum_{P>Q}\sum_{\ell>m}
\Lambda^Q_\ell\Lambda^P_m\sum_{s=0}^{\ell-m-1}
\bigg(t^{(\ell-s)N+Q}u^{mN+P}
\sum_{r=0}^{s-1}t^{(s-1-r)N}u^{rN}
\nonumber\\
+t^{mN+P}u^{(\ell-s)N+Q}\sum_{r=0}^{s-1}t^{rN}u^{(s-1-r)N}\bigg)
\nonumber\\
\fl=N(t^N-u^N)\sum_{P>Q}\sum_{\ell>m}
\Lambda^Q_\ell\Lambda^P_m\sum_{r=0}^{\ell-m-2}(\ell-m-1-r)
\nonumber\\
\times\,\big(t^{(\ell-1-r) N+Q}u^{(m+r)N+P}
+t^{(m+r)N+P}u^{(\ell-1-r)N+Q}\big).
\end{eqnarray}
Similarly we find
\begin{eqnarray}
\fl\alpha_2=N(t^N-u^N)\sum_{P>Q}\sum_{\ell<m}\Lambda^Q_\ell\Lambda^P_m 
\sum_{r=0}^{m-\ell-1}(m-\ell-r)\nonumber\\
\times\,\big(t^{(\ell+r)N+Q}u^{(m-r-1)N+P}
+t^{(m-r-1)N+P}u^{(\ell+r)N+Q}\big).
\end{eqnarray}
In the same way as decribed in detail in subsection {\ref{sec5.3}},
we can obtain the coefficients of $t^{\ell N+Q}u^{mN+P}$,
by first interchanging the order of the summations by moving the
sum over $r$ to the left, then changing the summation variables,
followed by interchanging the order of the summations in the
reverse order. Letting
$\alpha'_i\equiv{\alpha_i}/{N(t^N-u^N)}$, we find
\begin{eqnarray}
\fl\alpha'_1=\sum_{P>Q}\sum_{\ell=0}^{m_P-1}
\sum_{m=0}^{m_P-1}(t^{\ell N+Q}u^{mN+P}+u^{\ell N+Q}t^{mN+P})
\nonumber\\
\times\sum_{r=\max(0,m-\ell)}^{m}(\ell-m+r)
\Lambda^Q_{\ell+1+r}\Lambda^P_{m-r}, 
\label{alpha1p}\\ \cr
\fl{\alpha'_2}=\sum_{P>Q}\sum_{\ell=0}^{m_P-1}
\sum_{m=0}^{m_P-1}(t^{\ell N+Q}u^{mN+P}\!+\!u^{\ell N+Q}t^{mN+P})
\nonumber\\
\times\sum_{r=\max(0,\ell-m)}^{\ell}(m+1-\ell+r)
\Lambda^Q_{\ell-r}\Lambda^P_{m+r+1}. 
\label{alpha2p}\end{eqnarray}
From (\ref{Gtufi}) and (\ref{alpha12}), we may write
\begin{equation}
{\mathcal G}(t,u)=\psi'_1+\alpha'_1+\alpha'_2.
\end{equation}
From (\ref{alpha1p}) and (\ref{alpha2p}) we find that the coefficient
of $t^{\ell N+Q}u^{mN+P}$ is the same as the one of
$u^{\ell N+Q}t^{mN+P}$, and for $P>Q$ they are
\begin{eqnarray}
\fl{\mathcal G}_{\ell N+Q,mN+P}={\mathcal G}_{mN+P,\ell N+Q}=
\nonumber\\
\fl\sum_{r=\max(0,m-\ell)}^{m}(\ell-m+r)
\Lambda^Q_{\ell+1+r}\Lambda^P_{m-r}+
\sum_{r=\max(0,\ell-m)}^{\ell}(m+1-\ell+r)
\Lambda^Q_{\ell-r}\Lambda^P_{m+r+1}.
\end{eqnarray}
Replacing $r\to m-r$ in the first sum and $r\to\ell-r$ in the second,
we find
\begin{equation}
\fl{\mathcal G}_{\ell N+Q,mN+P}
=\sum_{r=0}^{\min(m,\ell)}\Big[(\ell-r)
\Lambda^Q_{\ell+m+1+r}\Lambda^P_{r} +(m+1-r)
\Lambda^Q_{r}\Lambda^P_{m+\ell+1-r}\Big],
\end{equation}
which is identical to (3.7) in \cite{AuYangPerk2010c}.
This complete the proof of our conjecture in \cite{AuYangPerk2010c}.

\section{Summary\label{sec6}}

Our previous work on the superintegrable chiral Potts model depended on
two conjectures. In our previous paper \cite{AuYangPerk2010b}, we made
the conjecture that the Serre relation are satisfied and in
\cite{AuYangPerk2010c} we conjectured a formula for the coefficients
of the two variable generating function ${\mathcal G}(t,u)$. Both these
conjectures were tested numerically. The validity of the Serre relations
is to be expected on the basis of the work of Davies \cite{Davies1990}
and one should be able to construct a proof extending work by Nishino
and Deguchi \cite{NiDe1,Degu1,NiDe2}.

In this paper we have managed to prove the conjecture for
${\mathcal G}(t,u)$ in \cite{AuYangPerk2010c}. In section 2, the
sum ${\mathcal V}(\{\mu_i\},\{\nu_i\})$ in (\ref{V}) was analyzed
using MacMahon's method described in \cite[chapter 11]{AndrewAskeyRoy}.
A theorem on symmetric funcions \cite{MacDonald,Sagan}, namely
(\ref{Sm}) was used next to relate the $L$-fold sum in (\ref{V}) to
an $m$-fold sums in (\ref{Sy}) by (\ref{Vf}). In section 3 we related
the sum ${\mathcal F}(t,u)$ in (\ref{Ftu}) by (\ref{FR}) to the sum
${\mathcal R}_{\ell,m}$ defined in (\ref{Rlm}), which was analyzed
further in (\ref{Rlm1}). In the remaining part of section 3, we got
rid of the equality signs in the summation and were able to arrive
at a simple formula for ${\mathcal F}(t,u)$ in (\ref{FR}) or
(\ref{Jk}), In section 4, we showed by induction the identity
(\ref{id4a}), allowing us to carry out the sums in (\ref{Jk}).
Again we used a well-known property of symmetric functions
\cite{MacDonald,Sagan} given in (\ref{poly}) to express
${\mathcal G}(t,u)$ in a much simpler form (\ref{Gtuf}). Finally,
in section 5, we analyzed the sum in (\ref{Gtuf}), and proved
the conjecture of \cite{AuYangPerk2010b}.


\end{document}